\documentclass[aip,amsmath,amssymb,reprint,superscriptaddress,floatfix]{revtex4-1}

\usepackage{graphicx}
\usepackage{floatrow}
\usepackage[caption=false,font={large}]{subfig}
\usepackage{bm}

\usepackage{amsmath}
\newcommand{\stkout}[1]{\ifmmode\text{\sout{\ensuremath{#1}}}\else\sout{#1}\fi}

\usepackage{amssymb}
\usepackage{color}
\usepackage{times}
\usepackage{float}
\usepackage[sort&compress]{natbib}

\usepackage[colorinlistoftodos,prependcaption,textsize=small]{todonotes}

\usepackage{hyperref}

\hypersetup{
  colorlinks=true,
citecolor=black,
linkcolor=black,
urlcolor=black
  }

\usepackage[T1]{fontenc}
\usepackage[utf8]{inputenc}

\newcommand{\kct}[1]{\textcolor{black}{#1}}

\begin{document}

\title{Optimization of escape kinetics by reflecting and resetting}
\author{Karol Capa{\l}a}
\email{k.capala@sanoscience.org}
\affiliation{Personal Health Data Science Group,\\
Sano - Centre for Computational Personalised Medicine,\\
Czarnowiejska 36, 30-054 Krak\'ow, Poland
}
\affiliation{Institute of Theoretical Physics
and Mark Kac Center for Complex Systems Research,
Faculty of Physics, Astronomy and Applied Computer Science,
Jagiellonian University, \L{}ojasiewicza 11, 30-348 Krak\'ow, Poland}

\author{Bart{\l}omiej Dybiec}
\email{bartlomiej.dybiec@uj.edu.pl}

\affiliation{Institute of Theoretical Physics
and Mark Kac Center for Complex Systems Research,
Faculty of Physics, Astronomy and Applied Computer Science,
Jagiellonian University, \L{}ojasiewicza 11, 30-348 Krak\'ow, Poland}


\date{\today}

\begin{abstract}
Stochastic restarting is a strategy of starting anew.
Incorporation of the resetting to the random walks can result in the decrease of the mean first passage time, due to the ability to limit unfavorably meandering, sub-optimal trajectories.
In the following manuscript, we examine how stochastic resetting influences escape dynamics from the $(-\infty,1)$ interval in the presence of the single-well power-law $|x|^\kappa$ potentials with $\kappa>0$.
Examination of the mean first passage time is complemented by the analysis of the coefficient of variation, which provides a robust and reliable indicator assessing efficiency of stochastic resetting.
The restrictive nature of resetting is compared to placing a reflective boundary in the system at hand.
In particular, for each potential, the position of the reflecting barrier giving the same mean first passage time as the optimal resetting rate is determined.
Finally, in addition to reflecting, we compare effectiveness of other resetting strategies with respect to optimization of the mean first passage time.
\end{abstract}


%
\maketitle

\setlength{\tabcolsep}{0pt}

\textbf{
The widespread occurrence of natural non-equilibrium systems operating under the influence of noise has captured significant attention in both theoretical and experimental investigations of dynamical stochastic systems.
Action of noise is responsible for spreading of trajectories.
The random spread of trajectories can be controlled by operation of deterministic forces leading to occurrence of noise induced effects resulting in ordered properties of noisy systems or optimal escape kinetics.
Another mechanism which can be used to control diffusive spread of particles is stochastic resetting.
Stochastic resetting is a motif that links diffusive motion and search strategies as starting anew can be used to optimize the time needed to find a target.
Interplay between deterministic forces and stochastic restarting can further facilitate the escape kinetics and increase the efficiency of search strategies.
}

%
%
\section{Introduction}
The noise driven escape from the semi-bounded domain is one of the archetypal problems in the theory of stochastic processes \cite{gardiner2009,redner2001,metzler2014firstpassage}.
It is often used within models of searching strategies \cite{shlesinger2009,viswanathan2011physics} which aims to find a target in the optimal way.
The minimization of the time of completion of the search is equivalent to reducing the time necessary to escape from the system as finding the target can be interpreted as the absorption.
One of the modifications which may lead to the acceleration of the escape is incorporation of the resetting protocol \cite{evans2020stochastic} because starting anew can limit wandering off in the wrong direction.
Main resetting schemes include return to a given point, e.g., the origin, after a deterministic (sharp resetting) \cite{pal2017first} or random (stochastic resetting) \cite{evans2011diffusion,nagar2016diffusion} amount of time or after crossing a given point $x_{sr}$ in space (spatial or first passage resetting) \cite{debruyne2020optimization,dahlenburg2021stochastic}.
Constructive action of all of these protocols originates in their ability to limit the space which particles can explore and elimination of wandering off.
Spatial resetting does it directly, while time-dependent resetting protocols perform it indirectly by limiting the time when particles can diffuse.

Stochastic resetting plays an important role not only in the first passage time problems, e.g., chemical kinetics \cite{reuveni2014role,rotbart2015michaelis}, search strategies \cite{lomholt2008levy,reynolds2009,viswanathan2011physics,palyulin2014levy}, but also in issues related to the existence and properties of (nonequilibrium) stationary states \cite{eule2016non}.
Starting anew changes properties of diffusive processes, thus it affects both first passage times and stationary states.
In single-well, power-law potential wells, stationary states do not exist if a particle can escape to infinity with non-vanishing probability.
This can happen (due too long jumps) for L\'evy noises \cite{jespersen1999,chechkin2002,chechkin2003,chechkin2004,dybiec2010d}  or (due to persistence) for the fractional Brownian noise \cite{guggenberger2021fractional,guggenberger2022absence}.
The stochastic resetting influences the problem of existence of stationary states because it efficiently eliminates long excursions and limits the spread of particles.
Therefore, it produces non-equilibrium stationary states in variety of situations, including free motion \cite{evans2011diffusion,nagar2016diffusion}, L\'evy flights \cite{stanislavsky2021optimal}, continuous time random walks \cite{mendez2021ctrw} or even motion in inverted potentials \cite{pal2015diffusion}.

The stochastic resetting is a protocol of starting anew \cite{evans2011diffusion,evans2020stochastic,gupta2022stochastic}.
The natural approach is to assume that restarts are triggered temporally, i.e., times of starting over are independent of the state of the system, e.g., position.
For example, resets can be performed periodically (sharp resetting) \cite{pal2017first} or with the constant rate (Poissonian resetting) \cite{evans2011diffusion}.
Time intervals between two consecutive resets can follow numerous distributions like:
exponential \cite{evans2011diffusion}, power-law \cite{nagar2016diffusion} and many others, e.g., Weibull, log-normal, log-logistic, Fr\'echet, see Refs.~\onlinecite{reuveni2016optimal,pal2017first}.
The endless options of temporal resetting protocols necessitate asking the question about generality and universality of restarting mechanisms.
Despite numerous underlying distributions relative fluctuations in first passage problems under restarts follow universal pattern  \cite{reuveni2016optimal,pal2017first}.
Therefore, the coefficient of variation not only displays universal properties but can be used to discriminate if starting anew can facilitate the escape kinetics.

The wandering off in the wrong direction can be eliminated by stochastic resetting, but it is not the only possible option.
The accessible space can be also limited in the more traditional way, i.e., by placing a reflecting boundary in the system.
\kct{Therefore, the reflection can play a key role in the optimization of the escape kinetics similarly to its pivotal role  in the stochastic description \cite{gadomski2001evolution,gadomski2003finite,weber2018entropy} of the nucleation\cite{kashchiev2000nucleation}.}
Analogously, to the spatial resetting, the barrier directly limits space, but instead of putting a particle back to a given point, e.g., the initial position, only prevents further motion in the given direction.
\kct{Consequently, on the one hand, it does not fully add the advantages of resetting, as it lacks the capability to start the motion anew, i.e., to relocate the particle to the restarting position.}
However, on the other hand, it is more restrictive than the so-called soft resetting \cite{xu2022stochastic}.

The model under study is described in the next section (Sec.~\ref{sec:model} Model).
Sec.~\ref{sec:results} (Results) analyzes properties of escape kinetics from the general single-well potential in presence of resetting and reflecting boundaries.
The manuscript is closed with Summary and Conclusions (Sec.~\ref{sec:summary}).

%
%
\section{Model\label{sec:model}}
The Langevin equation \cite{gardiner2009}
\begin{equation}
    \frac{d x}{dt}= -V'(x) + \xi(t),
    \label{eq:langevin}
\end{equation}
describes the noise driven motion of the overdamped particle in an external potential.
Here we assume that $V(x)$ is the general, single-well, power-law potential
\begin{equation}
    V(x)= |x|^\kappa\;\;\;\;\;(\kappa>0).
    \label{eq:potential}
\end{equation}
The random force $\xi(t)$ is approximated by the noise, which is modeled by the Gaussian white noise ($\langle \xi(t) \rangle = 0$ and $\langle \xi(t) \xi(s) \rangle=2 \sigma^2 \delta(t-s)$).
The parameter $\kappa$ in Eq.~(\ref{eq:potential}) controls the steepness of the potential.
In the limit of $\kappa\to\infty$, the single-well potential transforms into an infinite rectangular potential well with reflecting boundaries located at $\pm 1$.
\kct{Here, we restrict ourselves to the Gaussian white noise driving only.
Under action of the Gaussian white noise trajectories of the processes generated by Eq.~(\ref{eq:langevin}) are continuous and closed formulas for the mean first passage time are known.
Moreover, continuity of trajectories along with system symmetries simplifies the analysis of limiting cases because it allows for a transition between absorbing-absorbing and reflecting-absorbing setups, see the final part of Sec.~\ref{sec:results}.
}

We assume that the motion described by Eq.~(\ref{eq:langevin}) is limited to the $(-\infty,1)$ domain, because at \kct{$x=x_b=1$} the absorbing boundary is placed, while at $-\infty$ the external potential acts as the reflecting boundary.
Furthermore, the motion is affected by the so-called stochastic Poissonian resetting.
More precisely, we apply the restarting with a fixed rate $r$ ($r>0$).
The duration of time intervals $\tau$ between two consecutive resets follow the exponential distribution, $\phi(\tau)=r \exp{(-r \tau)}$, where $r$ is the restarting rate.
After each reset the motion is started anew from \kct{$x = x_0 \equiv x(0)$}.
The average time between two consecutive resets is equal to $1/r$.

First, we assess how Poissonian resetting \cite{evans2011diffusion,evans2011diffusion-jpa} can be used to optimize mean first passage time (MFPT) from the half-line $(-\infty,1)$ for various single-well potentials, see Eq.~(\ref{eq:potential}).
The MFPT  is the average of the first passage times
\begin{equation}
    \mathcal{T} =   \langle t_{\mathrm{fp}} \rangle =
 	\langle \min\{t : x(0)=x_0 \;\land\; x(t) \geqslant x_b  \} \rangle,
 	\label{eq:mfpt-eq}
\end{equation}
where $x_b=1$ is the location of the absorbing boundary.

In the absence of resetting, the $n$th moment $\mathcal{T}_n$ of first passage time ($\mathcal{T}_n = \langle t_{\mathrm{fp}}^n\rangle$ and $\mathcal{T}=\mathcal{T}_1=\langle t_{\mathrm{fp}} \rangle$) can be calculated from the recursive formula \cite{hanggi1990}
\begin{equation}
    -V'(x) \frac{\partial \mathcal{T}_n(x)}{\partial x} + \sigma^2 \frac{\partial^2 \mathcal{T}_n(x)}{\partial x^2}=-n \mathcal{T}_{n-1}(x).
    \label{eq:nMomentEq}
\end{equation}
In particular, for $n=1$, Eq.~(\ref{eq:2ndMomentEq}) gives the formula for the mean first passage time $\mathcal{T}=\mathcal{T}_1$, since the $0$th moment, due to normalization, is equal to 1, i.e., $\mathcal{T}_0=1$.
Eq.~(\ref{eq:nMomentEq}) is associated with the following boundary conditions: at the absorbing boundary $x_b$: $\mathcal{T}_n(x)\big|_{x=x_b}=0$, while for $V(x)=|x|^\kappa$ ($\kappa>0$) the reflecting boundary is formally placed at $-\infty$, therefore $\lim\limits_{x\to-\infty} \partial_x \mathcal{T}_n(x)=0$.

In the case of the escape from the interval restricted by the reflecting ($x_r$) and the absorbing boundary $x_b$ ($x_r<x_b$) for a particle starting its motion at $x$ the solution for the MFPT is given by, see Eq.~(5.5.23) in Ref.~\onlinecite{gardiner2009}
\begin{equation}
    \mathcal{T}(x|x_r,x_b)=\frac{1}{\sigma^2} \int_{x}^{x_b} \exp\left[ \frac{V(y)}{\sigma^2} \right]{d y} \int_{x_r}^{y}  \exp\left[ -\frac{V(z)}{\sigma^2} \right]d z.
    \label{eq:mfpt-ra-pot}
\end{equation}
Within current studies, we assume that $x_b=1$, $x_r \to -\infty$ and $x(0)=x=0$, \kct{unless stated otherwise}.
Eq.~(\ref{eq:mfpt-ra-pot}) gives the solution of Eq.~(\ref{eq:nMomentEq}) for $n=1$ with the appropriate boundary conditions \cite{gardiner2009}.

The advantage of stochastic resetting can be assessed by examination of the uninterrupted motion, i.e., motion without resets \cite{pal2017first}.
The coefficient of variation (CV) \cite{pal2017first}
\begin{equation}
    CV= \frac{\sigma(  t_{\mathrm{fp}})  }{ \langle t_{\mathrm{fp}} \rangle} = \frac{\sigma(  t_{\mathrm{fp}})}{\mathcal{T}}=\frac{\sqrt{\mathcal{T}_2-\mathcal{T}^2}}{\mathcal{T}} = \sqrt{  \frac{\mathcal{T}_2}{\mathcal{T}^2} -1},
    \label{eq:cv}
\end{equation}
which is the ratio between the standard deviation $\sigma(t_{\mathrm{fp}})$ of the first passage times and the mean first passage time $\mathcal{T}$ in the absence of stochastic resetting \cite{reuveni2016optimal,pal2017first,pal2019first},
can be used to find the domain where resetting can facilitate the escape kinetics \cite{reuveni2016optimal,ray2021resetting}.
The resetting is capable of accelerating the escape kinetics in the domain where $CV>1$.
Putting it differently, in the domain where $CV>1$, there is such an optimal resetting rate $r^*$ for which the mean first passage time is shorter than in the absence of resetting.
Moreover, for the motion under restarts $CV(r^*)=1$, see Refs.~\onlinecite{reuveni2016optimal,pal2017first}.

\begin{figure}[!h]
    \centering
  	   \includegraphics[angle=0,width=0.85\columnwidth]{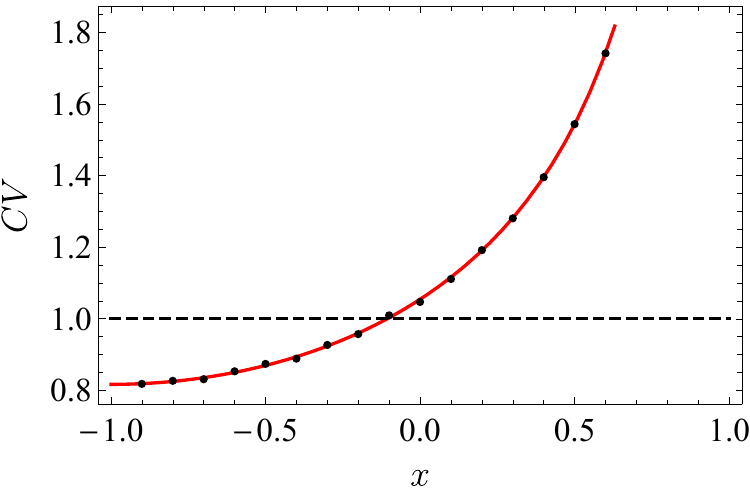}
  	   \caption{The coefficient of variation $CV(x)$ as a function of initial position $x$ for the escape from the interval restricted by the reflecting boundary at $x_r=-1$.
  	   Red solid line corresponds to the analytical solution given by Eq.~(\ref{eq:CVinf}) while black points represent results of numerical simulations.
  	   The black dashed line $CV=1$ defines $x$ which divides the plot into an area where resetting accelerates escape ($x>-0.1056$ where $CV>1$) and the area where it hinders ($x<-0.1056$ and $CV<1$) escape kinetics.}
    \label{fig:cvInf}
\end{figure}

Coefficient of variation, for the potential given by Eq.~(\ref{eq:potential}), can be expressed by elementary functions only for a few cases.
Formally for $\kappa=0$, Eq.~(\ref{eq:langevin}) describes the motion of a free particle.
In such a case, a particle certainly crosses the absorbing boundary, but  the first passage time distribution is given by the L\'evy-Smirnoff density \cite{redner2001,dybiec2009d} which is characterized by the diverging mean.
The mean first passage time can be rendered finite by stochastic resetting \cite{evans2011diffusion,evans2011diffusion-jpa,kusmierz2014firstorder}.
Results for $\kappa=1$, i.e., the motion in $V(x) \propto |x|$, are presented in Ref.~\onlinecite{singh2020resetting}.
Finally, another traceable case is $\kappa \to \infty$, which is equivalent to the motion of a free particle in the interval restricted by a reflecting boundary located at $x_r=-1$ ($\partial_x \mathcal{T}_n(x_r)=0$) and the absorbing one at $x_b=1$ ($ \mathcal{T}_n(x_b)=0$).
The formula for the MFPT can be obtained from Eq.~(\ref{eq:nMomentEq}) with $n=1$ and reads
\begin{equation}
   	  \mathcal{T}(x)= \frac{(3+x)(1-x)}{2\sigma^2}.
   \label{eq:MFPTinf}
\end{equation}
The second moment $\mathcal{T}_2$ can be also calculated from Eq.~(\ref{eq:nMomentEq}) with $n=2$, which transforms to
\begin{equation}
\sigma^2 \frac{d^2}{dx^2} \mathcal{T}_2(x)=-2\mathcal{T}(x).
\label{eq:2ndMomentEq}
\end{equation}
\kct{
Taking into account the boundary conditions ($\partial_x \mathcal{T}_2(x=-1)=0$ and $ \mathcal{T}_2(x=1)=0$, see Ref.~\onlinecite{gardiner2009}), the solution of Eq.~(\ref{eq:2ndMomentEq}) takes the following form }
\begin{eqnarray}
   	\mathcal{T}_2 &  = & \frac{1}{\sigma^4}\left(\frac{x^4}{12}+\frac{x^3}{3}-\frac{3 x^2}{2}-\frac{11 x}{3}+\frac{19}{4}\right) \nonumber \\
    	& = &  \frac{1}{12\sigma^4}\left(3+x\right)\left(1-x\right)(19-2x - x^2).
	\label{eq:T2}
\end{eqnarray}
\kct{From Eqs.~(\ref{eq:cv}) -- (\ref{eq:T2}), one gets}
\begin{equation}
 CV(x)=\sqrt{\frac{2}{3}} \sqrt{\frac{5+2x+x^2}{3-2x-x^2}}.
    \label{eq:CVinf}
\end{equation}
$CV$ given by Eq.~(\ref{eq:CVinf}) is the increasing function of the initial position $x$, on the $(-1,1)$ interval.
\kct{Please note, that initial position $x$ is the same as the position from which motion starts anew after each reset.}
As expected, $CV(x)$ does not depend on the noise intensity $\sigma$.
For $x=0$, $CV(0)\simeq 1.054$, therefore there should always be an optimal value of the resetting rate $r$ for which resetting shortens the escape time.
In fact, resetting should facilitate escape for every $x>-0.1056$, as it implies from the $CV$ criterion.
The dependence of $CV(x)$ on the initial position is shown in the Fig.~\ref{fig:cvInf}.
Intuitively, resetting can be beneficial if it moves a particle from a distant point to the vicinity of the target (absorbing boundary).
Therefore, resetting to a point which is distant from the absorbing boundary increases the average time needed to cross the boundary hindering the noise driven escape.
For the infinite rectangular potential well, resetting increases the MFPT (hinders escape kinetics) if the motion is restarted from $x<-0.1056$.
In further studies we set the noise intensity $\sigma$ to $\sigma=1$.

%
%
\section{Results\label{sec:results}}
\begin{figure}[!h]
    \centering
  	   \includegraphics[angle=0,width=0.85\columnwidth]{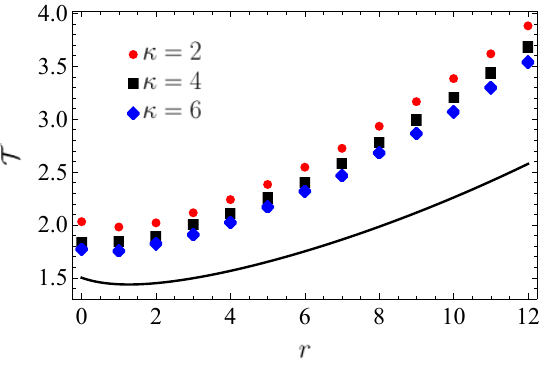}\\
    \includegraphics[angle=0,width=0.85\columnwidth]{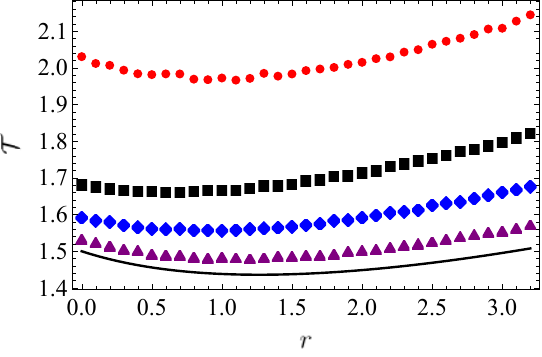}
  	   \caption{MFPT $\mathcal{T}(r)$ as a function of resetting rate $r$  for various exponents $\kappa$ characterizing steepness of the potential.
   Top panel shows MFPTs for $\kappa\in\{2,4,6\}$ and wide range of resetting rates $r$, while the bottom panel zooms the neighborhood of MFPTs' minima for wide range of $\kappa$, i.e., $\kappa=\{2,10,24,100\}$ (from top to bottom).
      Black solid line represents the analytical solution for $\kappa=\infty$, i.e., for the reflecting boundary located at $x_r=-1$.}
    \label{fig:resetting}
\end{figure}

Our study assumes that particle, which starts motion at $x=x_0=0$, moves in the potential  $V(x)=|x|^\kappa$ ($\kappa>0$) until it crosses the absorbing boundary located at $x_b=1$.
We begin our studies by inspecting the behavior of the MFPT as a function of the resetting rate $r$ for various values of the exponent $\kappa$, see Fig.~\ref{fig:resetting}.
Qualitatively, there is no significant difference between curves corresponding to various $\kappa$.
Initially, the MFPT decreases to a shallow minimum recorded for the optimal resetting rate $r^*$.
After reaching optimal value, further growth in $r$ leads to the increase in the MFPT.
For the fixed resetting rate $r$, the MFPT decreases with the increasing $\kappa$.
For larger resetting rates the difference becomes more evident, however the main difference between results with different exponents $\kappa$ is quantitative.

Top panel of Fig.~\ref{fig:resetting} shows the MFPT for selected values of exponents $\kappa$ ($\kappa\in\{2,4,6\}$), see Eq.~(\ref{eq:potential}), for a wide range of the resetting rates $r$.
It clearly demonstrates that after an initial decay, the MFPT starts to grow with the increasing resetting rate in the manner typical for Poissonian resetting.
The bottom panel displays the dependence of MFPT in the vicinity of its minima for a wide spectrum of exponents $\kappa$.
Therefore, it allows us to inspect how slowly the asymptotics, $\kappa \to \infty$, limit of the infinite rectangular potential well is reached\cite{dybiec2017levy}.
Moreover, it demonstrates that extraction of the optimal resetting rate $r^*$ can be hindered due to the shallow and broad minimum of MFPT.

\begin{figure}[!h]
    \centering
  	   \includegraphics[angle=0,width=0.85\columnwidth]{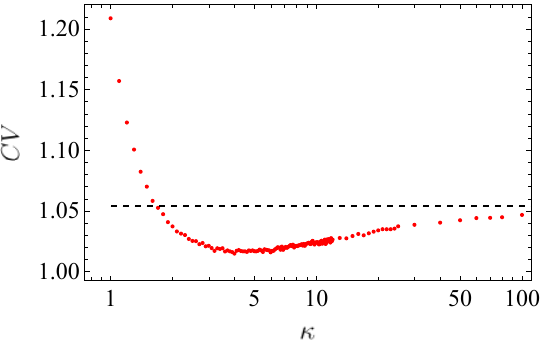}
  	   \caption{The coefficient of variation $CV(x=0)$ as a function of exponent $\kappa$. Black, dashed line corresponds to the asymptotic ($\kappa\to\infty$) value, i.e. $CV=1.054$, \kct{which is given by Eq.~(\ref{eq:CVinf}) with $x=0$, since for $\kappa\to\infty$ the reflecting boundary at $x_r=-1$ emerges and the escape problem is equivalent to the escape from $(-1,1)$ interval restricted by reflecting and absorbing boundaries}.}
    \label{fig:cv}
\end{figure}

Influence of the resetting on the MFPT, see Eqs.~(\ref{eq:langevin}) and (\ref{eq:potential}), can be also indirectly examined using the coefficient of variation $CV$.
Fig.~\ref{fig:cv} shows the coefficient of variation for the initial position $x=0$, i.e.,  $CV(x=0)$, as a function of the exponent $\kappa$.
For all studied exponents $\kappa$, $CV>1$, therefore there is always an optimal value of the resetting rate $r^*$ which minimizes the MFPT.
Nevertheless, the $CV$ exhibits interesting non-monotonic behavior.
After the rapid initial decay, $CV$ reaches its minimum.
Afterward it slowly increases to reach the asymptotic value $CV \to 1.054$ for $\kappa \to \infty$, which corresponds to the infinite rectangular potential well, see Eq.~(\ref{eq:CVinf}) and Ref.~\onlinecite{dybiec2017levy},
\kct{i.e., to the escape of a free particle from the interval $(-1,1)$ restricted by reflecting ($x_r=-1$) and absorbing ($x_b=1$) boundaries.}

The dependence of the $CV$ on the steepness exponent $\kappa$ can be intuitively explained.
For very small $\kappa$, e.g., $\kappa<2$, the deterministic force is not efficient in pulling particles back to the origin.
They can broadly spread and quite easily move towards $-\infty$.
For large $\kappa$ the potential becomes steeper resulting in strong repulsive force.
The strong repelling force reduces the probability of finding particles at distant points on the left side of the origin.
Consequently, the role played by stochastic resetting is decreased.
Nevertheless, for all exponents $\kappa$, the stochastic restarting of the motion can reduce the MFPT.

\begin{figure}[!h]
    \centering
  	   \includegraphics[angle=0,width=0.85\columnwidth]{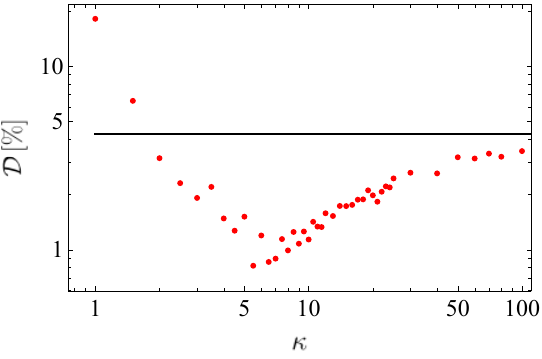}
  	   \caption{The MFPT drop $\mathcal{D}=\left[ 1-\mathcal{T}_{r=r^*}(\kappa)/\mathcal{T}_{r=0}(\kappa) \right] \times 100\;\%$, see Eq.~(\ref{eq:drop}), as a function of the potential steepness $\kappa$. The solid line shows $\mathcal{D}(\kappa\to\infty)$ asymptotics, see Eq.~(\ref{eq:dropinf}).	 
  	   }
    \label{fig:mfptgain}
\end{figure}

Since the $CV$ analysis showed that resetting hastens the escape kinetics, the natural next step is to quantify the effectiveness of stochastic restarting.
Fig.~\ref{fig:mfptgain} shows the numerically estimated MFPT drop $\mathcal{D}$ which measures facilitation of the escape kinetics due to resetting
\begin{equation}   
 \mathcal{D}(\kappa)=\left[ 1-\frac{\mathcal{T}_{r=r^*}(\kappa)}{\mathcal{T}_{r=0}(\kappa)} \right] \times 100 \;\%,
 \label{eq:drop}
\end{equation}
where $\mathcal{T}_{r=r^*}(\kappa)=\min_r\{\mathcal{T}_r(\kappa)\}$ is the minimal MFPT corresponding to optimal resetting rate $r^*$, while $\mathcal{T}_{r=0}(\kappa)$ stands for the MFPT for the uninterrupted motion.
Large values of the drop indicate significant facilitation of the escape kinetics.
Fluctuations in $\mathcal{D}$ are due errors in estimation of MFPT and difficulties in precise extracting of the optimal resetting rate $r^*$, see Fig.~\ref{fig:resetting}.
One may see that the MFPT drop shows initial decay, as with the increasing $\kappa$, penetration of the outer ($ x<-1$) area becomes more difficult.
For $\kappa \approx 7$, drop $\mathcal{D}(\kappa)$ attains minimum value and starts to increase.
For large $\kappa$, distant points ($x<-1$) are practically not explored, but at the same time the restoring force at $-1<x<1$ becomes weak.
Finally, for $\kappa\to\infty$ the motions resemble free motion restricted by reflecting and absorbing boundaries.
Therefore, in the limit of $\kappa\to\infty$, the MFPT drop $\mathcal{D}$ attains the infinite rectangular potential well limit, see below.
Moreover, as it was already mentioned above, typically the MFPT drop is not significant, due to the confining nature of the potential with $\kappa>2$.

\begin{figure}
    \centering
    \includegraphics[angle=0,width=0.85\columnwidth]{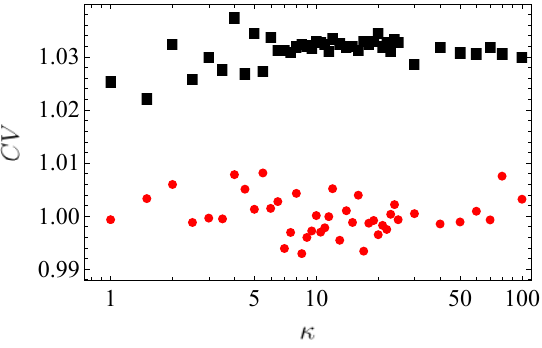}
    \caption{Coefficient of variation for the optimal resetting rates  $r^*$ (red bullets), see Fig.~\ref{fig:resetting}, and the fixed resetting rate $r=3.2$ (black squares) as the function of the exponent $\kappa$, see Eq.~(\ref{eq:potential}).}
    \label{fig:cvres}
\end{figure}

Fig.~\ref{fig:cvres} display the value of the coefficient of variation at the optimal resetting rate $r^*$, i.e., $CV(0|r^*)$ (red bullets), and $CV(0|r=3.2)$ (black squares).
The coefficient of variation at $r^*$ as predicted by general considerations \cite{reuveni2016optimal,pal2017first} regarding properties of fluctuations in first passage times is equal to unity.
At the same time, for $r=3.2$, coefficient of variation is visibly larger than for $r=r^*$ and as such deviates from unity showing a possible space for further optimization.

\begin{figure}[!h]
    \centering
  	   \includegraphics[angle=0,width=0.85\columnwidth]{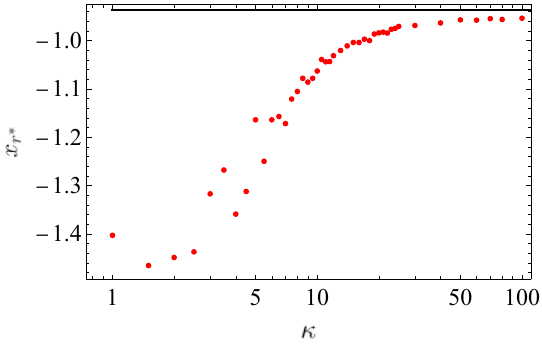}\\
  	   \caption{The position of reflecting boundary $x_{r^*}(\kappa)$ which gives the same MFPT as the optimal resetting rate as a function of the steepness exponent $\kappa$.
   The solid line shows $x_{r^*}(\kappa\to\infty)=-0.936$ asymptotics.
  	   }
    \label{fig:xr}
\end{figure}

In the next step we compare stochastic resetting with the other protocol of eliminating meandering trajectories.
Resetting can efficiently eliminate long trajectories exploring distant (towards $-\infty$) points but spread of trajectories can be also eliminated by introduction of the reflecting boundary.
Reflecting boundary can be placed at any point $x_r$ to the left of the initial position, i.e., $x_r \in (-\infty,0)$.
Nevertheless, we place the reflecting boundary in a very special place.
For each $\mathcal{T}_{r=r^*}(\kappa)$ we have found the location of the reflecting boundary $x_{r^*}$, which assures the same MFPT as the optimal resetting with the resetting rate $r^*$.
The reflecting boundary acts in a similar manner like resetting, because it prevents exploration of points $x<x_r$.
However, in contrast to resetting, it does not start the motion anew.
\kct{The position of the reflecting boundary $x_{r^*}$ corresponding to $\mathcal{T}_{r=r^*}(\kappa)$ was computed using Eq.~(\ref{eq:mfpt-ra-pot}).
Specifically, we determined $x_{r^*}$ such that Eq.~(\ref{eq:mfpt-ra-pot}) yields the same MFPT as the optimal resetting in the absence of a reflecting boundary.}
Fig.~\ref{fig:xr} shows the location $x_{r^*}(\kappa)$ of the reflecting barrier assuring the same MFPT as optimal resetting
\begin{equation}
	\mathcal{T}(x=0,x_{r^*},x_b=1) = \mathcal{T}_{r=r^*}(\kappa).
	\label{eq:xrcondition}
\end{equation}
Analogously like in Fig.~\ref{fig:mfptgain} fluctuations in $x_{r^*}$ are due to difficulties in precise extracting of the optimal resetting rate $r^*$ as minima of MFPT curves are shallow and wide, see bottom panel of Fig.~\ref{fig:resetting}.
The $\kappa \to \infty$ limit corresponds to the placing of the reflecting boundary within the infinite rectangular potential well, as for $\kappa\to\infty$ the particle motion reduces to the free motion restricted by reflecting ($x_r=-1$) and absorbing ($x_b=1$) boundaries.
Therefore, the limiting $\kappa\to\infty$ case can be studied more rigorously, see below.
With the increasing $\kappa$, $x_{r^*}$ moves steadily to the right.
Finally, it saturates around $x_{r^*} \approx -0.936$ which corresponds to $\kappa\to\infty$ limit, i.e., to the stochastic resetting of free motion in the presence of the reflecting boundary at $x_r=-1$.
\kct{
Further shift of the reflecting boundary to the right (beyound $x_{r^*}$) can subsequently decrease the value of the MFPT below $\mathcal{T}_{r=r^*}(\kappa)$.
The minimal MFPT is reached when the reflecting boundary is placed at the initial position.
Thus, reflecting can easily outperform temporal stochastic resetting.
}

\kct{
The motion under stochastic resetting and the uninterrupted (restricted by the reflecting boundary) motion bear fundamental differences.
Nevertheless, both processes, despite being very different, can display the  MFPT equivalence.
In other words, for $r^*$, it is possible to find an effective escape problem from a bounded interval  $(x_{r^*},1)$ leading to the same MFPT.
The location of the reflecting boundary $x_{r^*}$ moves towards the origin as $\kappa$ increases, see Fig.~\ref{fig:xr}, in response to changes in the potential.
The exponent $\kappa$ controls the steepens of the $|x|^\kappa$ potential, which in the $\kappa\to\infty$ limit reduces to the infinite rectangular potential well.
As $\kappa$ increases, the external potential increasingly restricts the exploration of the region $x\ll 0$ and decreases the MFPT.
For this reason, $\mathcal{T}_{r=r^*}(\kappa)$ is a decreasing function of $\kappa$.
Similarly, $\mathcal{T}(x=0,x_{r},x_b=1)$ is a decreasing function of $x_r$ ($x_r<0$).
As a result, the condition on $x_{r^*}$, i.e., $\mathcal{T}(x=0,x_{r^*},x_b=1) = \mathcal{T}_{r=r^*}(\kappa)$ demands $x_{r^*}$ to move toward the origin until it reaches the value corresponding to the $\kappa\to\infty$ limit.
}

The formula for the MFPT under stochastic resetting  \cite{pal2019first}  from the interval $(-L,L)$ restricted by two absorbing boundaries
\begin{equation}
\mathcal{T}(x_0) = \frac{1}{r} \left[ \frac{  \sinh  \frac{2L}{\sqrt{  \sigma^2/r}}   }{	\sinh \frac{L-x_0}{\sqrt{ \sigma^2/r}}   +\sinh   \frac{x_0+L}{\sqrt{  \sigma^2/r}}  } -1 \right]
	\label{eq:mfptreset}
\end{equation}
can be adapted to the studied (reflecting-absorbing) setup by the following substitution $L\to 2$, $\sigma^2 \to 1$ and $x_0 \to 1$.
The formula~(\ref{eq:mfptreset}) can be transformed to the escape from the interval restricted by the reflecting boundary at $x=-1$ and the absorbing boundary at $x_b=1$, because escape from such a setup (under action of the Gaussian white noise) is equivalent to the escape from two times wider interval restricted by two absorbing boundaries and the appropriately shifted initial position $x_0$.
From Eq.~(\ref{eq:mfptreset}) with $L = 2$, $\sigma^2 = 1$ and $x_0 = 1$ it is possible to calculate
$\mathcal{T}_{r=r^*} \approx 1.436$ which is recorded for optimal resetting rate $r^* \approx  1.266$.
The MFPT for the non-interrupted motion can be calculated either from Eq.~(\ref{eq:MFPTinf}) or as the $r\to 0$ limit of Eq.~(\ref{eq:mfptreset}) resulting in $\mathcal{T}_{r=0}=\frac{3}{2}$.
Using these values, one gets
\begin{equation}
\mathcal{D}(\kappa\to\infty) \approx 4.26\;\%,
\label{eq:dropinf}
\end{equation}
showing that subsequent increase in $\kappa$, see Fig.~\ref{fig:mfptgain}, can further optimize the MFPT, because the minimal drop is well below the asymptotic value.
Moreover, the MFPT drop slowly grows with the increase in $\kappa$.

Finally, we would like to comment on the other possibility of the MFPT optimization.
In addition to temporal resetting, there are other resetting scenarios called a position (spatial) dependent resetting \cite{dahlenburg2021stochastic}, i.e., restarting is performed every time when a particle reaches a given point $x_{sr}$.
Such a resetting scheme can be even more efficient than reflecting, because it not only prevents exploration of points $x<x_{sr}$ but also immediately brings a particle back to the restarting position $x=0$.
Therefore, a particle, due to resetting, can reach an intermediate point (origin), which is needed to be passed through, faster than under reflection.

\begin{figure}[!h]
	\centering
	\includegraphics[angle=0,width=0.85\columnwidth]{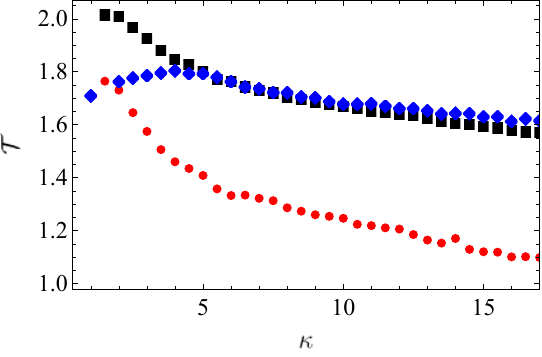}\\
 \includegraphics[angle=0,width=0.85\columnwidth]{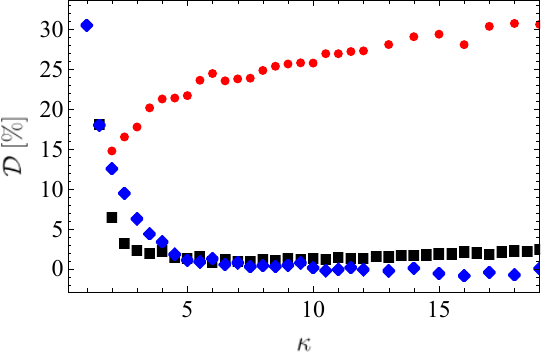}
	\caption{The top panel shows MFPT for the escape with reflecting boundary located in the optimal $x_{r^*}$ (black squares), with the spatial resetting boundary $x_{sr}$ located at $x_{r^*}$ (red dots) and for the fixed position of the resetting boundary placed at $x_{sr}=-1.5$ (blue rhombuses).
 The middle panel shows corresponding coefficients of variation, see Eq.~(\ref{eq:cv}).
 The bottom panel presents corresponding drops $\mathcal{D}$, see Eq.~(\ref{eq:drop}), for various optimization strategies.}
	\label{fig:spatial}
\end{figure}

Figure~\ref{fig:spatial} compares various optimization protocols.
Top panel shows MFPT values, while the bottom panel displays MFPT drops $\mathcal{D}$.
The largest values of the MFPT are recorded for the uninterrupted motion, which is chosen as the reference point.
The stochastic Poissonian resetting, which with respect to the MFPT is equivalent to placing a reflecting boundary in an appropriate selected point $x_{r^*}$ (see Eq.~(\ref{eq:xrcondition}) and Fig.~\ref{fig:xr}), slightly decrease the MFPT (black squares).
Nevertheless, for $\kappa>3$, the MFPT is practically equal to the MFPT for the uninterrupted motion.
Thus, for the figure's clarity we do not show results for the uninterrupted ($r=0$) motion.
If the reflection at $x_{r^*}$ is replaced by the spatial resetting at $x_{sr}=x_{r^*}$ the MFPT declines significantly (red circles).
Additional blue rhombuses correspond to spatial resetting at the fixed ($\kappa$ independent) position $x_{sr}=-1.5$.
On the one hand, spatial resetting at $x_{sr}=-1.5$ with $\kappa<5$ outperforms stochastic resetting, nevertheless it is not as efficient as the spatial resetting at $x_{sr}=x_{r^*}$, see Eq.~(\ref{eq:xrcondition}).
On the other hand, in order to trigger the spatial resetting at $x_{sr}=-1.5$ it is not required to perform any calculations.
The efficiency of the spatial resetting can be improved by shifting $x_{sr}$ to the right.
Finally, placing the reflecting boundary at $x_r=0$, i.e., at the starting position, is even more efficient than any type of resetting to $x=0$, because points $x<0$ are not visited at all.
In such a case in the limit of $\kappa\to\infty$ the MFPT is equal to $\frac{1}{2}$ as it can be deducted from the modified version of Eq.~(\ref{eq:MFPTinf}) (it is enough to replace 3 by 1 in the numerator and substitute $x=0$).

The MFPT drops $\mathcal{D}$ in the bottom panel of Fig.~\ref{fig:spatial} are calculated in respect to the uninterrupted ($r=0$) motion.
From drops it is clearly visible that spatial resetting at $x_{sr}=x_{r^*}$ outperforms other studied types of resetting.
Moreover, it demonstrates that for $\kappa<5$ spatial resetting at $x_{sr}=-1.5$ is a plausible optimization strategy.
Finally, for large exponents ($\kappa>5$) the external potential efficiently bounds particle's trajectories making spatial resetting at $x_{sr}=-1.5$ less efficient than uninterrupted motion or the Poissonian resetting.
In such a case, the MFPT drop falls slightly below 0 as the spatial resetting for $x_{sr}=-1.5$ produces larger MFPTs than uninterrupted motion.

%
%
\section{Summary and conclusions \label{sec:summary}}

Resetting and introduction of the reflecting boundaries are two methods capable of reducing the mean first passage time.
Both of them \kct{achieve this} by weakening the ability of a particle to explore parts of the space distant from the absorbing barrier.
However, they differ in the details.
A reflecting boundary limits the domain of motion by preventing a particle from visiting points outside the boundary while stochastic resetting reduces the time of space exploration, leading to removal of random parts of the space.
Spatial resetting works similarly but additionally to the exclusion of the (fixed) part of the domain of motion, it also brings the particle immediately to the starting position.

In this paper, we explored influence of these three methods of the escape enhancement on a motion in the semi-bounded interval in the presence of external potential of the $|x|^{\kappa}$ form with $\kappa>0$.
By direct examination, and by the analysis of coefficient of variation, we showed that the Poissonian resetting is always beneficial in such a system.
Increasing $\kappa$ initially leads to a decay in the MFPT because it becomes harder for particles to penetrate the outer area ($x<-1$).
The drop $\mathcal{D}(\kappa)$ reaches its minimum value around $\kappa\approx7$, and then starts to increase.
In the limit $\kappa\to\infty$ results tend to one for the free motion with reflecting and absorbing boundaries.
Moreover, since resetting and presence of the reflecting boundary facilitate escape by restricting available space, there is such a location of the reflecting boundary $x_{r^*}$ that gives the same mean first passage time as the temporal Poissonian resetting with the optimal resetting rate $r^*$.
The location of the absorbing boundary moves toward the origin with the increasing steepness exponent $\kappa$ and tends to a fixed value.
Spatial resetting, which restarts motion every time when a given point is crossed, outperforms both temporal stochastic resetting and reflecting of trajectories.

%
%
\section*{Acknowledgements}

We gratefully acknowledge Poland’s high-performance computing infrastructure PLGrid (HPC Centers: ACK Cyfronet AGH) for providing computer facilities and support within computational grant no. PLG/2023/016175.
The research for this publication has been supported by a grant from the Priority Research Area DigiWorld under the Strategic Programme Excellence Initiative at Jagiellonian University.
This research has been supported by the European Union’s Horizon $2020$ research and innovation programme under grant agreement Sano No $857533$.
This publication is supported by Sano project carried out within the International Research Agendas programme of the Foundation for Polish Science, co-financed by the European Union under the European Regional Development Fund.

\section*{Data availability}
The data  (generated randomly using the model presented in the paper) that support the findings of this study are available from the corresponding author (KC) upon reasonable request.

%
%

\section*{References}
\def\url#1{}

\end{document}